# Unification of Gravitational and Electromagnetic Fields in Riemannian Geometry

*Yi-Fang Chang*

*Department of Physics, Yunnan University, Kunming 650091, China*

(E-mail: yifangchang1030@hotmail.com)

**Abstract**: The gravitational field and the source-free electromagnetic field can be unified preliminarily by the equations $R^i_{klm} = \kappa T^i_{klm}{}^*$ in the Riemannian geometry, both are contractions of im and ik, respectively. If $R^i_{klm} = \kappa T^i_{klm}{}^*$ =constant, so it will be equivalent to the Yang's gravitational equations $R_{km;l} - R_{kl;m} = 0$, which include $R_{lm} = 0$. From $R_{lm} = 0$ we can obtain the Lorentz equations of motion, the first system and second source-free system of Maxwell field equations. This unification can be included in the gauge theory, and the unified gauge group is SL(2,C)×U(1)=GL(2,C), which is just the same as the gauge group of the Riemannian manifold.

In order to unify the gravitational and electromagnetic fields, Einstein, et al., applied various geometrical structures and theoretical schemes, for example, the gauge invariance geometry, the five-dimensional space-time, the projective theory [1], the affine field and the bivector fields, etc. The non-symmetric field "is the logically simplest relativistic field theory which is at all possible" [2,3]. Hlavaty made the summary and development [4]. Johnson expounded systematically his viewpoints [5]. Gaffney [6] and Moffat, Boal [7] obtained some conclusions and solutions. The unified field theories of more than four-dimensions were summarized [8]. Madore [9] proposed a modification of the traditional formulation of Kaluza-Klein theory in which the internal structure is described by a noncommutative geometry based on a semisimple algebra. The classical theory of Yang-Mills fields and of Dirac fermions is developed in the resulting geometry. The generalized connection is written down which should describe the unification of the Yang-Mills fields with gravity. Moreover, the unification is connected with the string model [10,11] and the grand unification theory [11], and the supersymmetry [12] and supergravity, etc.

But, because the unified field theory is a very complex and difficult question, further approach should be made still. Recently, the unification of quantum field theory and general relativity as a fundamental goal of modern physics is discussed. Solanki, et al., studied new couplings between electromagnetism and gravity [13]. In grand unified theories with large numbers of fields, renormalization effects modify the scale at which quantum gravity becomes strong. Calmet, et al. shown that these effects from gravity can be larger than the two-loop corrections considered in renormalization group analyses of unification [14].

We introduced a principle of equivalence for the electromagnetic field: A non-inertial system with an acceleration is equivalent to a certain electromagnetic field, in which the ratio of charge to mass is the same [15]. From this principle an electromagnetic general relativity theory (GRT) can be derived, whose formulations are completely analogous to Einstein's GRT. In the electromagnetic case, the field is regarded as a type of curved space-time for charged bodies, where space-time is separated into many layers, the curvatures of which are different for different ratios of charge to mass. In a general case, electrodynamics can be obtained from this theory. Shch the gravitational and electromagnetic fields can be unified. But its high-order approximation will deviate from the present electromagnetic theory. Therefore, we discussed the four possible tests for this theory and some notable problems. Finally, the most universal principle of extended equivalence and the extended GRT are proposed [16].

Einstein, et al., always believed that the unified field should have no sources [2,1]. We think that the gravitational field and the source-free electromagnetic field can be unified preliminarily only the Riemannian geometry [17].

In the Riemannian geometry there are the equations [18]:

$$R^i_{klm} = \kappa T^i_{klm} *. \tag{1}$$

Here 
$$R^i_{klm} = \frac{\partial \Gamma^i_{km}}{\partial x_l} - \frac{\partial \Gamma^i_{kl}}{\partial x_m} + \Gamma^i_{nl}\Gamma^n_{km} - \Gamma^i_{nm}\Gamma^n_{kl}. \tag{2}$$

In this case $\Gamma^i_{kl}$ are the Christoffel symbols $\left\{ \begin{matrix} i \\ kl \end{matrix} \right\}$. Lanczos pointed out also that the unification should turn to $R_{iklm}$, but he applied $H_{ikl}$ [19].

By the contraction of im, the Einstein gravitational field equations:

$$R_{kl} = \frac{\partial \Gamma^i_{ki}}{\partial x_l} - \frac{\partial \Gamma^i_{kl}}{\partial x_i} + \Gamma^i_{nl}\Gamma^n_{ki} - \Gamma^i_{ni}\Gamma^n_{kl} = \kappa T_{kl}* = \kappa(T_{kl} - \frac{1}{2}g_{kl}T), \tag{3}$$

are derived. The corresponding geodesic equation of motion is

$$\frac{du^\mu}{ds} + \Gamma^\mu_{\alpha\beta} u^\alpha u^\beta = 0. \tag{4}$$

By the contraction of ik, Eqs.(2) become

$$R_{lm} = R^k_{klm} = \frac{\partial \Gamma^k_{km}}{\partial x_l} - \frac{\partial \Gamma^k_{kl}}{\partial x_m} + \Gamma^k_{nl}\Gamma^n_{km} - \Gamma^k_{nm}\Gamma^n_{kl}. \tag{5}$$

In the Riemannian geometry ik is antisymmetrical, so

$$R_{lm} = 0, \tag{6}$$

i.e., 
$$\frac{\partial \Gamma^k_{km}}{\partial x_l} - \frac{\partial \Gamma^k_{kl}}{\partial x_m} = -\Gamma^k_{nl}\Gamma^n_{km} + \Gamma^k_{nm}\Gamma^n_{kl}. \tag{7}$$

It is analogous to the Lorentz equation of motion:

$$\frac{\partial A_m}{\partial x_l} - \frac{\partial A_l}{\partial x_m} = F_{lm} = \frac{mw_l}{ev^m}. \tag{8}$$

Assume that the corresponding relation

$$\Gamma^k_{km} = y_m \Rightarrow A_m, \tag{9}$$

hold, then the left of Eq.(7) becomes the electromagnetic tensor $\frac{\partial A_m}{\partial x_l} - \frac{\partial A_l}{\partial x_m} = F_{lm}$. Because $p_l = mv_l + eA_l$, so $mv_l/e$ corresponds to $A_l$ and $\Gamma^k_{kl}$. $\Gamma^k_{nl}$ are divided into two parts: $\Gamma^k_{kl}$ (k=n) and $\Gamma^k_{n'l}$ (k≠n'), so the first term of the right of Eq.(7) is

$$-\Gamma^k_{nl}\Gamma^n_{km} = -(\Gamma^k_{kl}+\Gamma^k_{n'l})\Gamma^n_{km} \Rightarrow -(mv_l/e)\Gamma^n_{km} - \Gamma^k_{n'l}\Gamma^n_{km}. \tag{10}$$

Based on Eq.(4), $\Gamma^n_{km} = -dv^n/ds(v^k v^m)$. We take this part of k=n, then the first term of Eq.(10) is just

$$(-\frac{mv_l}{e})(-\frac{dv^k}{dsv^k v^m}) = (\frac{mdv_l/ds}{j^m})\frac{dv^k}{v^k}\frac{v_l}{dv_l} = \frac{mw_l}{j^m}. \tag{11}$$

Therefore, $R_{lm}=0$ includes the Lorentz equation $mw_l = F_{lm}j^m$. In the Riemannian geometry $\Gamma^k_{nl} \neq 0$ shows the curved space-time, in which a part corresponds to the Lorentz electromagnetic force.

Eq.(6) is differentiated, and we introduce an additional condition

$$\frac{\partial(\Gamma^\alpha_{\beta\mu}\Gamma^\beta_{\alpha\nu})}{\partial x_\nu} = 0, \tag{12}$$

which is analogous to the Lorentz condition $\frac{\partial A_i}{\partial x_i} = 0$. Such

$$\frac{\partial R_{lm}}{\partial x_m} = \frac{\partial F_{lm}}{\partial x_m} = 0. \tag{13}$$

This is namely the second system of Maxwell equations with $j_l = 0$.

According to $\frac{\partial \Gamma^k_{km}}{\partial x_l} - \frac{\partial \Gamma^k_{kl}}{\partial x_m} \Rightarrow F_{lm}$, the first system of Maxwell equations

$$\frac{\partial F_{lm}}{\partial x_k} + \frac{\partial F_{mk}}{\partial x_l} + \frac{\partial F_{kl}}{\partial x_m} = 0, \tag{14}$$

may be obtained directly, and they are analogous completely to the Bianchi identities.

In the gauge theory [20] the gauge field strength is:

$$f_{lm}^{k} = \frac{\partial b_{l}^{k}}{\partial x_{m}} - \frac{\partial b_{m}^{k}}{\partial x_{l}} - b_{l}^{i} b_{m}^{j} c_{lj}^{k}, \tag{15}$$

where $b_{l}^{k}$ is a gauge field potential. They have included the electromagnetic field of G=U(1) and the Yang-Mills field of G=SU(2). Yang supposed

$$b_{l}^{(ik)} = \Gamma_{kl}^{i}, \quad f_{lm}^{(ik)} = -R_{klm}^{i}, \tag{16}$$

so the Riemannian geometry is obtained. Of course, the gravitational field equations (3) and the electromagnetic field equations (6) can be obtained, since both are particular cases of the Riemannian geometry. The gauge field potentials $b_{l}^{k} = \Gamma_{kl}^{k}$ corresponds to the electromagnetic potentials $A_{l}$, the structure constant is

$$c_{(\lambda\sigma)(\eta\rho)}^{(kk)} = \delta_{\sigma\eta} \delta_{k\lambda} \delta_{k\rho} - \delta_{\lambda\rho} \delta_{k\eta} \delta_{k\sigma}. \tag{17}$$

So the gauge field strength

$$-f_{lm}^{k} = -(\frac{\partial \Gamma_{kl}^{k}}{\partial x_{m}} - \frac{\partial \Gamma_{km}^{k}}{\partial x_{l}} - \Gamma_{\sigma l}^{\lambda} \Gamma_{\rho m}^{\eta} c_{(\lambda\sigma)(\eta\rho)}^{(kk)}) = \frac{\partial \Gamma_{km}^{k}}{\partial x_{l}} - \frac{\partial \Gamma_{kl}^{k}}{\partial x_{m}} + \Gamma_{\sigma l}^{k} \Gamma_{km}^{\sigma} - \Gamma_{kl}^{\lambda} \Gamma_{\lambda m}^{k} = R_{klm}^{k}, \tag{18}$$

correspond to the electromagnetic field strength $F_{lm}$. When $c_{ij}^{k} = 0$, and the group is U(1), etc., the electromagnetic force is zero, $R_{lm} = F_{lm} = 0$.

We know that the gravitational field equations possess two distinct invariances: GL(4,R) invariance of Einstein under coordinate transformations, and SL(2,C) gauge invariance of Weyl [21]. Carneli [22] thought that the group SL(2,C) replaced the group SU(2) of the Yang-Mills field equations, and the gravitational field equations in free space can be derived. The gauge group G of electromagnetic field is U(1). Therefore the unified field should possess the gauge group SL(2,C)×U(1)=GL(2,C) [23]. Yang derived that parallel displacement defines a gauge field with G being GL(n) in the Riemannian manifold [20]. Such the unification of the gravitational and electromagnetic fields may be derived preliminarily in the general Riemannian geometry. Solanki, et al., considered a metric-affine gauge theory of gravity in which torsion couples nonminimally to the electromagnetic field [13].

In the Riemannian geometry, by the contraction the Bianchi identities become

$$R_{klm;l}^{i} + R_{km;l} - R_{kl;m} = 0. \tag{19}$$

So for pure space the Yang's gravitational equations

$$R_{km;l} - R_{kl;m} = 0. \tag{20}$$

are equivalent to the free equations

$$R_{klm;l}^{i} = 0, \tag{21}$$

i.e., the Riemannian curvature $R_{klm}^{i}$ =constant. For general case Eqs.(21) should be extended to

Eqs.(1). By the contraction of im or ik, Eqs.(21) turn to

$$R_{kl;i} = 0, \quad T_{kl;i}* = 0, \tag{22}$$

$$R_{lm;i} = 0, \text{ and } R_{lm} \equiv 0. \tag{6}$$

For a particular case Eqs.(22) are

$$R_{kl} = 0. \tag{23}$$

This is namely the Einstein gravitational field equations in free space. Therefore Eqs.(20) include Eqs.(22), (23), (6) and $F_{lm;m} = 0$. Of course, the static spherical symmetric metric derived from Eqs.(20) include the Schwarzschild solution derived from Eqs.(23).

In fact, for various geometries in which Eqs.(2) hold, such as the non-symmetrical field [2-5] and Brans-Dicke theory, etc. [24,25], no matter what the concrete contents of various connection coefficients $\Gamma^i_{kl}$ are, we can always obtain the Lorentz equations of motion, the first system and second source-free system of Maxwell field equations, if only we suppose that Eqs.(9), (11), (12) hold, and use similar contraction of im, and let $R_{lm} = 0$. Of course, the Riemannian geometry is the simplest case in all. We think that the unification of the gravitational and electromagnetic fields by the geometric methods should be researched continuously.

**References**
1. O.Veblen and B.Hoffmann, Phys.Rev. 36,810(1930).
2. A.Einstein, The Meaning of Relativity (Fifth edition). Princeton.1955.
3. D.H.Boal and J.W.Moffat, Phys.Rev. D11,2026(1975).
4. V.Hlavaty, Geometry of Einstain's Unified Field Theory. P.Noordhoff Ltd. 1957.
5. C.R.Johnson, Phys.Rev. D4,295; 318; 3555(1971); D5,282; 1916(1972); D7,2825; 2838; D8, 1645(1973).
6. G.W.Gaffney, Phys.Rev. D10,374(1974).
7. J.W.Moffat and D.H.Boal, Phys.Rev. D11,1375(1975).
8. V.D.Sabbata and E.Schmutzer, Unified Field Theories of More Than 4 Dimensions. 1983.
9. J.Madore, Phys.Rev. D41,3709(1990).
10. M.Chemtob, Phys.Rev. D56,2323(1997).
11. K.Dimopoulos, Phys.Rev. D57,4629(1998).
12. M.A.Luty, Phys.Rev.Lett. 89,141801(2002).
13. S.K.Solanki, O.Preuss, M.P.Haugan, et al., Phys.Rev. D69,062001(2004).
14. X.Calmet, S.D.Hsu and D.Reeb, Phys.Rev.Lett. 101,171802(2008).
15. Yi-Fang Chang, Matter Regularity. 3,75(2003).
16. Yi-Fang Chang, Galilean Electrodynamics. 16,91(2005).
17. Yi-Fang Chang, New Research of Particle Physics and Relativity. Yunnan Science and Technology Press. 1989. p184-216. Phys.Abst.93,1371(1990).
18. C.W.Kilmister and D.J.Newman, Proc.Cam.Phil.Soc. 57,851(1961).
19. C.Lanczos, Rev.Mod.Phys. 29,337(1957).

20. C.N.Yang, Phys.Rev.Lett. 33,445(1974).
21. C.J.Isham, A.Salam and J.Strathdee, Nuovo Cimento. 5,969(1972).
22. M.Carmeli, Nucl.Phys. B38,621(1972).
23. V.G.Kadyshevsky, et al., Phys.Lett. 15,182(1965).
24. C.Brans and R.H.Dicke, Phys.Rev. 124,925(1961).
25. C.W.Misner, S.K.Thorne and J.A.Wheeler, Gravitation. W.H.Freeman and Company. 1973.